\documentclass[%
reprint,
amsmath,amssymb,
aip,
]{revtex4-2}
\usepackage{bm}
\usepackage{graphicx}   
\usepackage{subfigure}
\usepackage{dcolumn}
\usepackage{xcolor}

\begin{document}
\title{Hot-carrier thermophotovoltaic systems} 
\author{Kartika N. Nimje}
\author{Maxime Giteau}
\email[]{maxime.giteau@icfo.eu}
\author{Georgia T. Papadakis}
\email[]{georgia.papadakis@icfo.eu}
\affiliation{ICFO - Institut de Ciencies Fotoniques, The Barcelona Institute of
Science and Technology, Castelldefels (Barcelona) 08860, Spain}
\date{\today}

\begin{abstract}
{A thermophotovoltaic (TPV) energy converter harnesses thermal photons emitted by a hot body and converts them to electricity. When the radiative heat exchange between the emitter and PV cell is spectrally monochromatic, the TPV system can approach the Carnot thermodynamic efficiency limit. Nonetheless, this occurs at the expense of vanishing extracted electrical power density. Conversely, a spectrally broadband radiative heat exchange between the emitter and the cell yields maximal TPV power density at the expense of low efficiency. By leveraging hot-carriers as a means to mitigate thermalization losses within the cell, we demonstrate that one can alleviate this trade-off between power density and efficiency. Via detailed balance analysis, we show analytically that one can reach near-Carnot conversion efficiencies close to the maximum power point, which is unattainable with conventional TPV systems. We derive analytical relations between intrinsic device parameters and performance metrics, which serve as design rules for hot-carrier-based TPV systems.}
\end{abstract}


\maketitle

\section{Introduction}
\par{Thermophotovoltaic (TPV) energy conversion recycles heat into electricity via a system consisting of a thermal emitter coupled to a photovoltaic (PV) cell. The emitter, which is externally heated, radiates thermal photons toward a PV cell, which, in turn, converts them into electricity \cite{harder_theoretical_2003, chubb_fundamentals_2007, datas_steady_2013, karalis_squeezing_2016, papadakis_thermodynamics_2021}. In recent years, various TPV systems have been realized, reporting record-high conversion efficiencies \cite{lapotin_thermophotovoltaic_2022,omair_ultraefficient_2019,datas_thermophotovoltaic_2021}. The remarkable progress in TPV performance has attracted applications in low-grade waste heat recovery \cite{zhao_high-performance_2017}, intermittent power generation \cite{giteau_thermodynamic_2023}, and heat-to-energy conversion at ultra-high temperatures \cite{tervo_efficient_2022, lapotin_thermophotovoltaic_2022}.}

\par{Nonetheless, unless sophisticated design takes place, for example, by placing the thermal emitter in the thermal near-field of the cell \cite{mittapally_near-field_2021, song_modeling_2022}, reported high TPV efficiencies come at the cost of considerably compromised extracted electrical power densities. In fact, the extracted electrical power density and the conversion efficiency are by nature competing performance TPV metrics. The extracted electrical power density, termed $P_\mathrm{el}$ henceforth, expresses the power consumed by a load connected to the TPV module. The conversion efficiency, defined as $\eta=P_\mathrm{el}/P_\mathrm{exc}$, expresses the portion of the radiative heat exchange between the emitter and the cell, $P_\mathrm{exc}$, that is converted into electricity by the cell. Various recent approaches have explored ways to simultaneously increase both the conversion efficiency and the power density of TPV devices, for example, by operating at very high temperatures \cite{tervo_efficient_2022, lapotin_thermophotovoltaic_2022}, optimizing photon recycling \cite{fan_near-perfect_2020, lee_air-bridge_2022, lim_enhanced_2023}, enhancing the absorption view factor \cite{lopez_thermophotovoltaic_2023}, and leveraging the thermal near-field \cite{pendry_radiative_1999, papadakis_thermodynamics_2021, lucchesi_near-field_2021, pascale_perspective_2023}. Nevertheless, the trade-off between power density and conversion efficiency remains a major challenge in TPV design.}

\par{In an ideal TPV device, in the limit of spectrally monochromatic radiative heat exchange between the emitter and cell, one can approach the Carnot thermodynamic efficiency limit, denoted as $\eta_\mathrm{C}=1-T_\mathrm{C}/T_\mathrm{H}$, where $T_\mathrm{C}$ is the temperature of the cell and $T_\mathrm{H}$ is the temperature of the emitter. Approaching the Carnot efficiency, however, leads to negligible power output. On the other hand, broadening the spectrum of radiative heat exchange improves $P_\mathrm{el}$ but comes at the cost of lower $\eta$. In \cite{giteau_thermodynamic_2023}, we presented the explicit relation between the fundamental limits of $P_\mathrm{el}$ and $\eta$ for radiative energy conversion. In the present work, we propose a strategy to overcome the power-efficiency trade-off in single-junction TPVs.}

\par{The notion of maximum extracted electrical power density in a single-junction TPV cell is analogous to the Shockley-Queisser limit \cite{shockley_detailed_2004} for single-junction solar cells. Multiple strategies have been proposed to overcome the Shockley-Queisser limit in solar PVs \cite{green_third_2006}. The most widespread and record-holding approach is to employ multi-junction solar cells, where each cell converts a portion of the solar spectrum, thereby improving the current-voltage characteristics of the cell \cite{vos_detailed_1980, geisz_six-junction_2020, green_solar_2023}. This concept has also been explored for TPV systems \cite{datas_optimum_2015,lapotin_thermophotovoltaic_2022}. In this case, the TPV power density is increased, and as a consequence, the TPV efficiency follows.}

\par{Here, we explore an alternative approach to improve $P_\mathrm{el}$ and $\eta$ simultaneously by leveraging hot-carriers, in analogy to their implementation in hot-carrier solar cells (HCSC) \cite{ross_efficiency_1982}. In HCSC systems, the excess heat generated through absorption of high-energy photons improves the voltage characteristics of the cell, upon the extraction of electron-hole pairs prior to their thermalization. This can be achieved by extracting the hot carriers using energy-selective contacts (ESC)\cite{wurfel_solar_1997, marti_thermodynamics_2022}, and results in a theoretical efficiency limit of 85$\%$ for zero bandgap under full concentration \cite{green_third_2006}. Although the experimental efficiency of HCSC remains currently limited due to strong thermalization rates in most materials, promising avenues have been identified \cite{konig_non-equilibrium_2020, zhang_review_2021}, using ultrathin \cite{giteau_identification_2020} and low-dimensional absorbers, such as quantum wells \cite{rosenwaks_hot-carrier_1993,hirst_enhanced_2014,nguyen_quantitative_2018, makhfudz_enhancement_2022}, nanowires \cite{fast_hot-carrier_2021} and dots \cite{harada_hot-carrier_2019}, or hybrid hot-carrier multi-junction devices \cite{giteau_hot-carrier_2022-1, giteau_hot-carrier_2022}.}

\par{We note that recent studies that have also considered hot carriers in TPV systems have primarily focused on the selection of materials comprising the emitter and cell, and only pertain to near-field operation \cite{st-gelais_hot_2017, wang_hot_2023}. By contrast, in this work, we identify performance limits of hot-carrier-based TPV (HCTPV) systems in the far-field. Using a detailed balance formalism \cite{shockley_detailed_2004}, we compute the extracted electrical power density and efficiency for any combination of semiconductor bandgap and emitter temperature, both analytically and numerically. As such, this work presents a general framework describing the physics of HCTPV systems.}

\begin{figure}
    \centering
    \includegraphics[width=\linewidth]{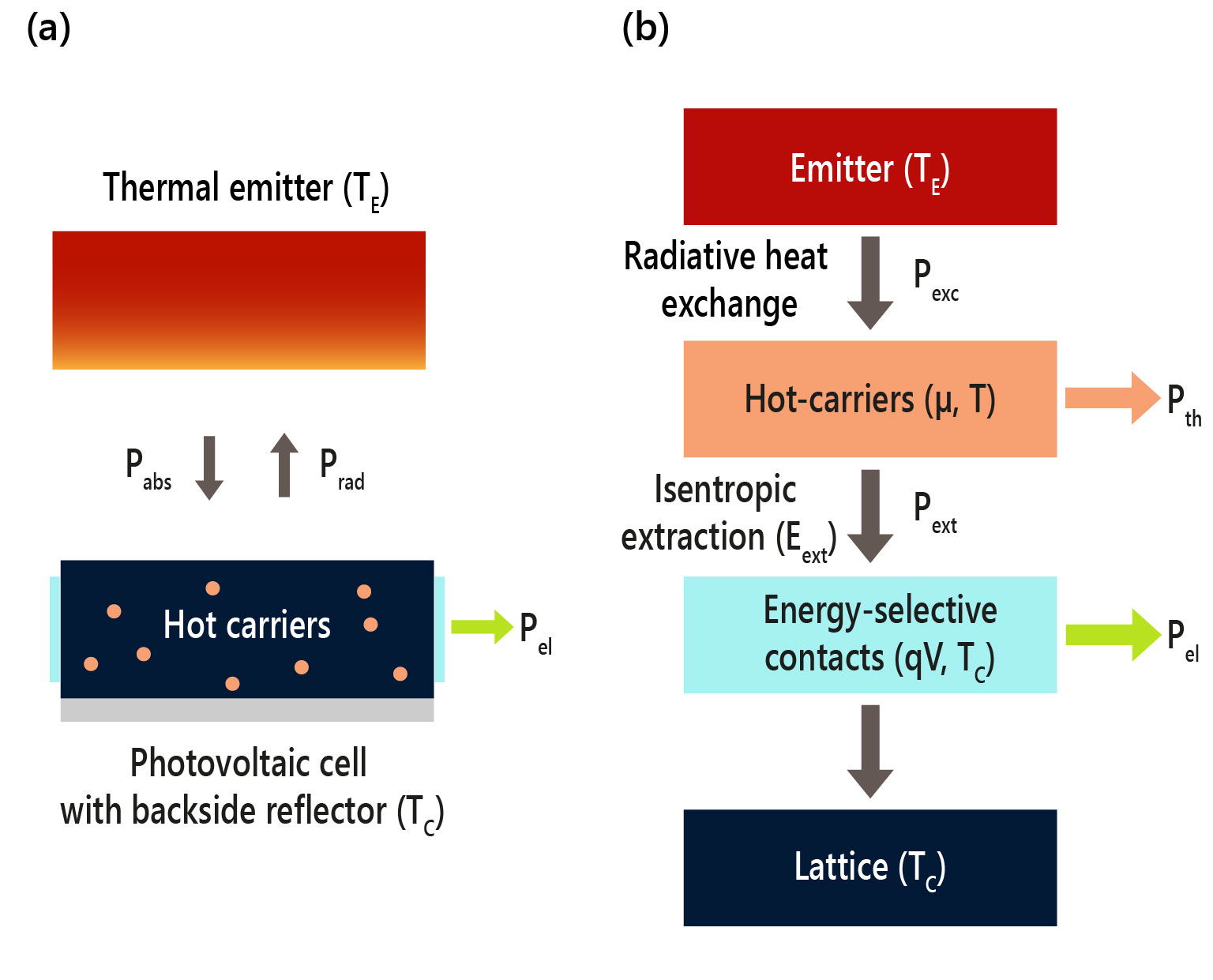}
    \caption{(a) Schematic of a hot-carrier thermophotovoltaic (HC-TPV) system: A blackbody emitter at temperature $T_\mathrm{E}$ exchanges thermal radiation with a PV cell. Perfect photon recycling is considered, represented visually with a perfect backside reflector (grey) at temperature $T_\mathrm{C}$. The radiative power densities, denoted by $P_\mathrm{abs}$ and $P_\mathrm{rad}$, are associated with the absorption and emission of photons by the PV cell respectively. Each absorbed photon generates high-energy \emph{hot} carrier. A HCTPV device extracts the energy of these hot-carriers as electrical power density, $P_\mathrm{el}$, via the selective contacts shown in cyan. (b) Thermodynamic representation of the HCTPV device using isentropic extraction at energy $E_\mathrm{ext}$.}
    \label{fig:Figure 1}
\end{figure}

\par{We demonstrate that a HCTPV system offers a significantly higher maximum power density as compared to a standard TPV. At the maximum power point, efficiency is also very high as compared to a standard TPV system. Finally, we show that, at $V=V_\mathrm{OC}$, a HCTPV always reaches the Carnot thermodynamic efficiency limit, a feature unattainable in standard TPV operation. As such, a single-junction HCTPV offers a physical approach to reaching thermodynamic limits of radiative heat engines, as described in [10] \cite{giteau_thermodynamic_2023}.

\section{Theoretical formalism}
\par{Our model is analogous to the one introduced by Ross and Nozik for HCSCs \cite{ross_efficiency_1982}, where the sun is replaced by a local blackbody source at temperature $T_\mathrm{E}$. We consider that the PV cell has a temperature $T_\mathrm{C}$ and a bandgap $E_\mathrm{g}$, as shown in Fig. \ref{fig:Figure 1}(a). To simplify the analysis, both the emitter and cell are modeled as semi-infinite parallel slabs to eliminate any radiation leakage, ensuring a unity view factor \cite{burger_present_2020}. In addition, we consider that the absorptivity of the cell is unity above the bandgap and that below-bandgap radiative heat exchange between the emitter and the cell is completely suppressed. This can be achieved in practice with an ideal backside reflector ensuring perfect sub-bandgap photon recycling \cite{omair_ultraefficient_2019} as shown in Fig. \ref{fig:Figure 1}(a). We also do not consider near-field contributions to the radiative heat exchange between the emitter and the cell. The analysis is conducted in the radiative limit, for which non-radiative losses in the PV cell are neglected. Finally, we consider that each absorbed photon generates a single electron-hole pair with the same energy.}

\begin{figure*}
\includegraphics[width=\linewidth]{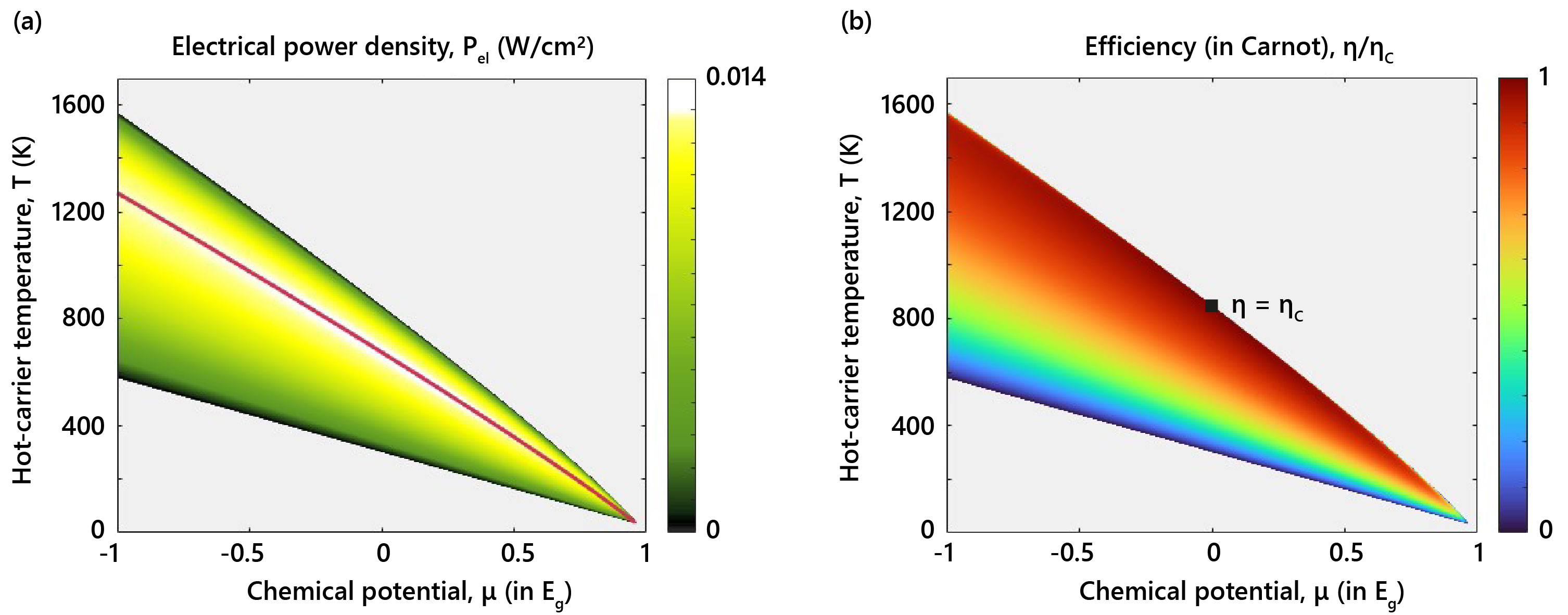}
\caption{\label{fig2} (a) Electrical output power density and, (b) efficiency, analytically derived up to a second-order Maxwell-Boltzmann approximation, as a function of normalised chemical potential and temperature of the hot-carriers at $T_\mathrm{E} = 850  \text{K}$, $T_\mathrm{C} =  300 \text{K}$ and $E_\mathrm{g} = 0.74  \text{eV}$. The red curve in (a) represents the temperature as a function of chemical potential for which the power density is maximum. The black point in (b) indicates the open-circuit voltage, that is, $\left(\mu, T\right) = \left(0, T_\mathrm{E}\right)$ at which the efficiency reaches Carnot. The gray regions correspond to $P_\mathrm{el} <0$ (not relevant for TPV operation). }
\label{fig:Figure 2}
\end{figure*}

\par{Since only photons with energy $\varepsilon \geq E_\mathrm{g}$ are exchanged, the particle flux (in $\mathrm{m^{-2}s^{-1}}$) absorbed by the cell can then be expressed as: 
\begin{align}
    \label{eq: defn_j_abs}
    N_\mathrm{abs} &= \frac{1}{4 \mathrm{c}^2 \pi^2 \hbar^3} \int_{E_\mathrm{g}}^{\infty} \frac{\varepsilon^2}{\exp{{\left(\dfrac{\varepsilon}{k_\mathrm{B} T_\mathrm{E}}\right)}} - 1}\, d\varepsilon
\end{align}
\noindent where $\mathrm{c}$ is the speed of light in vacuum, $k_\mathrm{B}$ is the Boltzmann constant, and $\hbar$ is the reduced Planck constant. We assume that carrier-carrier scattering occurs sufficiently fast so that photo-generated carriers follow a Fermi-Dirac distribution characterized by a quasi-Fermi level splitting $\mu$ (hereafter referred to as \emph{chemical potential}) and a temperature $T$. Thus, the photon flux radiated by the cell follows the generalized Planck law \cite{wurfel_chemical_1982}:
\begin{align}
    \label{eq: defn_j_rad}
    N_\mathrm{rad}\left(\mu, T \right) &= \frac{1}{4 \mathrm{c}^2 \pi^2 \hbar^3} \int_{E_g}^{\infty} \frac{\varepsilon^2}{\exp{{\left(\dfrac{\varepsilon - \mu}{k_\mathrm{B} T}\right)}} - 1}\, d\varepsilon.
\end{align}}
\par{Under continuous illumination, steady-state is achieved when the rate of photo-generation is balanced with the combined rate of recombination and extraction of carriers. Based on the principle of detailed balance introduced by Shockley and Queisser \cite{shockley_detailed_2004}, the extracted current density, $J_\mathrm{ext}$ (in $\mathrm{Am^{-2}}$), can be written as:
\begin{align}
    \label{eq: extracted_current_density}
    J_\mathrm{ext}\left( \mu, T \right) &= \mathrm{q}\left[N_\mathrm{abs} - N_\mathrm{rad}\left( \mu, T \right)\right]
\end{align}
\noindent where $\mathrm{q}$ is the electron charge. In the absence of hot carriers, in other words, when carriers dissipate their excess kinetic energy through thermalization, upon reaching thermal equilibrium with the lattice, it holds that $T = T_\mathrm{C}$. In that case,  and assuming ideal carrier transport, the chemical potential $\mu$ is equal to $\mathrm{q}V$, where $V$ is the applied voltage \cite{wurfel_solar_1997}. By solving Eq. (\ref{eq: extracted_current_density}) as a function of $\mu$ using Eqs. (\ref{eq: defn_j_abs}) and (\ref{eq: defn_j_rad}), we obtain the well-known $J_\mathrm{ext}-V$ characteristic of a PV cell. By contrast, in a hot-carrier PV cell, the carriers are not in thermal equilibrium with the lattice ($T \neq T_\mathrm{C}$), and the carrier population is a function of both $\mu$ and $T$.}

\par{We introduce the quantity $P_\mathrm{th}$, which represents the energy lost in thermalization, upon the excitation of carriers. The extracted power density of carriers equals $P_\mathrm{ext}=P_\mathrm{exc}-P_\mathrm{th}$, where $P_\mathrm{exc}$ is the radiatively exchanged power density between the emitter and the cell, expressed as:
\begin{align}
    \label{eq: extracted_power_density}
      P_\mathrm{exc}\left(\mu, T\right) &= P_\mathrm{abs} - P_\mathrm{rad}\left( \mu, T \right),
\end{align}
\noindent where $P_\mathrm{abs}$ and $P_\mathrm{rad}$ are the power densities absorbed and radiated by the cell, respectively. These can be obtained by multiplying the integrand in Eqs. \ref{eq: defn_j_abs}-\ref{eq: defn_j_rad} by $\varepsilon$.}

\par{To harness the energy of hot carriers with minimal loss, following \cite{wurfel_solar_1997}, we consider isentropic extraction, expressed through:
\begin{equation}
    \label{eq: defn_hotcarrier_qV}
            \mathrm{q}V = \mu\frac{T_\mathrm{C}}{T} + \left(1 - \frac{T_\mathrm{C}}{T} \right) E_\mathrm{ext}.
\end{equation}
\noindent In Eq. (\ref{eq: defn_hotcarrier_qV}), the extraction energy, termed $E_\mathrm{ext}$, is the difference between the energy levels of extraction for electrons and holes, and satisfies $P_\mathrm{ext} = J_\mathrm{ext} E_\mathrm{ext}/\mathrm{q}$ \cite{giteau_detailed_2019}. As a reference, we note that, in the limit $T = T_\mathrm{C}$ for a conventional PV cell, from Eq. (\ref{eq: defn_hotcarrier_qV}), we indeed retrieve the standard expression $\mu = \mathrm{q} V$.} 

\par{The electrical power density, $P_\mathrm{el}$, is defined as $P_\mathrm{el} = J_\mathrm{ext}V$ (Fig. \ref{fig:Figure 1}(b)). Consequently, the conversion efficiency in the general case of a HCTPV is given by 
\begin{equation}
    \label{eq: defn_hotcarrier_efficiency}
            \eta = \frac{P_\mathrm{el}}{P_\mathrm{ext}+P_\mathrm{th}}.
\end{equation}}

\par{In the limit where thermalization is completely suppressed and carriers remain at temperature $T$, which is the case for HCTPV considered here, $P_\mathrm{th}=0$, and $P_\mathrm{ext}=P_\mathrm{exc}$. From Eq. (\ref{eq: defn_hotcarrier_efficiency}), $\eta=P_\mathrm{el}/P_\mathrm{exc}$ as introduced above.}

\section{Results and discussion}
\begin{figure*}
\includegraphics[width=\linewidth]{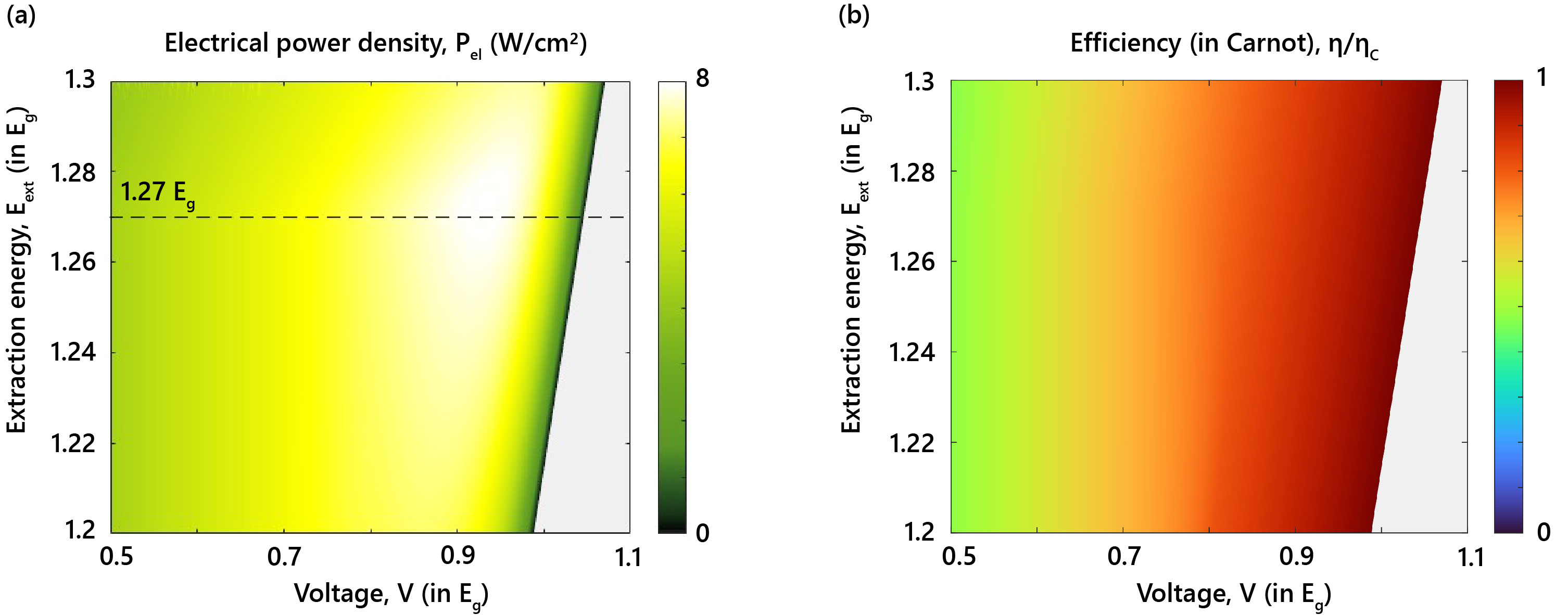}
\caption{\label{fig3} (a) Electrical output power density and (b) efficiency calculated as a function of the device parameters (extraction energy and voltage) for $T_\mathrm{E} = 1700  \text{K}$, $T_\mathrm{C} =  300 \text{K}$ and $E_\mathrm{g} = 0.74  \text{eV}$. The dashed line in (a) indicates the extraction energy that maximizes the power density. The gray regions correspond to $P_\mathrm{el} <0$ (LED operation).}
\label{fig:Figure 3}
\end{figure*}

\par{The analysis considered here pertains to an idealized HCTPV system where thermalization is completely suppressed. The temperature of the carriers is $T$, whereas the temperature of the lattice is $T_\mathrm{C}$ (Fig. \ref{fig:Figure 1}(b)). From Eq. (\ref{eq: extracted_power_density}) and the expressions of $P_\mathrm{ext}$ and $P_\mathrm{el}$, the conversion efficiency Eq. (\ref{eq: defn_hotcarrier_efficiency}) reduces to: 
\begin{equation}
\label{eq: defn_efficiency_hot}
    \eta = \frac{\mathrm{q} V}{E_\mathrm{{ext}}}.
\end{equation}}

\par{From Eq. (\ref{eq: defn_efficiency_hot}), the maximum efficiency is obtained at $V=V_\mathrm{oc}$, for which, here, we derive general analytical expressions for the efficiency and open-circuit voltage of the HCTPV system. At $V=V_\mathrm{oc}$, $J_\mathrm{{ext}} = 0$ and $P_\mathrm{{ext}} = 0$. Solving Eqs. (\ref{eq: extracted_current_density})-(\ref{eq: extracted_power_density}) at $V=V_\mathrm{oc}$ yields $\mu = 0$ and $T = T_\mathrm{E}$. The open circuit voltage Eq. (\ref{eq: defn_hotcarrier_qV}) and the corresponding efficiency Eq. (\ref{eq: defn_efficiency_hot}) are, respectively, given by:
\begin{equation}
    \label{eq: voc_second_order_purelyhot}
    V_\mathrm{{oc}} = \left(1-\frac{T_\mathrm{C}}{T_\mathrm{E}}\right) \frac{E_\mathrm{{ext}}}{\mathrm{q}} 
\end{equation}
\begin{equation}
\label{eq: efficiency_hot_voc}
    \eta_\mathrm{oc} = 1 - \frac{T_\mathrm{C}}{T_\mathrm{E}} = \eta_\mathrm{C}
\end{equation}
Eq. (\ref{eq: efficiency_hot_voc}) demonstrates that an ideal HCTPV system always reaches the Carnot thermodynamic efficiency limit ($n_\mathrm{C}$). This is not the case for conventional TPV systems, unless the bandwidth of the radiative heat exchange between the emitter and the cell approaches zero \cite{datas_thermophotovoltaic_2021}. In a conventional TPV system, $\eta$ is also maximal at $V_\mathrm{oc}$, nonetheless this maximized efficiency does not approach $\eta_\mathrm{C}$. At the same time, near $V_\mathrm{oc}$, the power density of a conventional TPV system vanishes, whereas in a HCTPV this is no longer true. As demonstrated in Fig. \ref{fig:Figure 4} (a) of the following analysis, in a HCTPV system, one approaches maximal power density at near-Carnot efficiencies.} 

\par{As a reference, in \cite{giteau_thermodynamic_2023}, by considering reciprocal radiative heat exchange between a blackbody coupled to a Carnot engine, termed the endoreversible approximation, we analyzed the thermodynamic performance bounds of radiative heat engines. The analysis of a HCTPV system outlined above reduces to the endoreversible approximation in the limit of $E_\mathrm{G}=\mu=0$. As a result, a zero-bandgap HCTPV system approaches the thermodynamic bound of radiative heat engines. By contrast, approaching this bound with conventional TPV cells would require an infinite number of junctions (multicolor limit) \cite{green_third_2006}.}

\par{Next, we identify the optimal internal parameters (hot carrier temperature, $T$, and chemical potential, $\mu$, in Fig. \ref{fig:Figure 2}) as well as external device parameters (extraction energy, $E_\mathrm{ext}$, and voltage, $V$, in Fig. \ref{fig:Figure 3}) for optimal HCTPV performance. We note that, in the limit $\left(E_g-\mu\right)/ k_\mathrm{B} T \gg 1$, in other words, within the Maxwell-Boltzmann approximation, it is possible to derive analytical expressions. In the Appendix, Eqs. (A9), (A11), we present the relation between $\mu$ and $T$ for optimal power density and efficiency, respectively, up to a second-order approximation.} 

\begin{figure*}
    \centering
     \includegraphics[width=\linewidth]{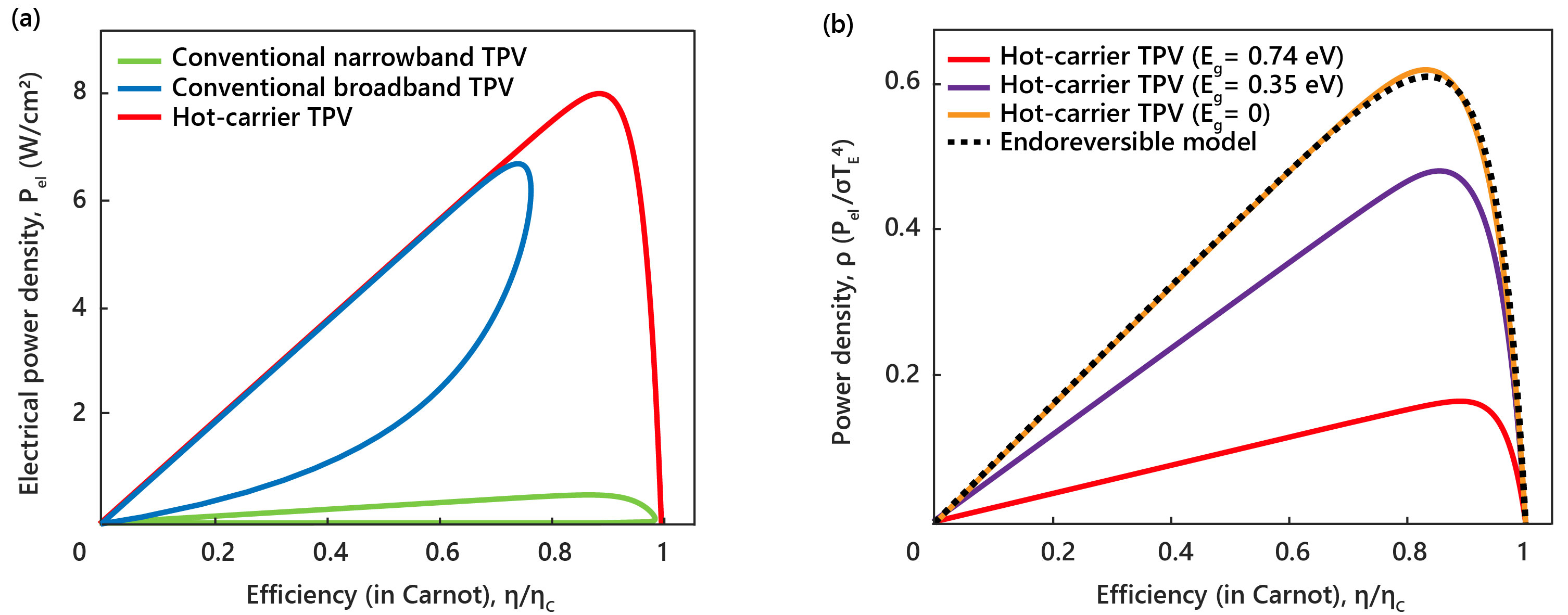}
     \caption{(a) Electrical power density ($P_\mathrm{el}$) versus conversion efficiency in terms of the Carnot limit ($\eta/\eta_\mathrm{C}$) for $(E_\mathrm{g} = 0.74  \text{eV})$. Red line: HC-TPV at power-maximizing extraction energy $E_\mathrm{ext} = 1.27 E_\mathrm{g}$. Blue line: conventional broadband TPV. Green line: conventional narrowband TPV (emission bandwidth of 15 meV above the bandgap). (b) Power density normalised to $\sigma T_\mathrm{E}^4$ ($\rho$) versus efficiency in terms of the Carnot limit ($\eta/\eta_\mathrm{C}$) for $E_\mathrm{g} = 0.74  \text{eV}$, $E_\mathrm{g} = 0.35  \text{eV})$ and a zero-bandgap cell (the optimal extraction energies are $E_\mathrm{ext} = 0.94 \text{eV}$, $0.59 \text{eV}$ and $0.41 \text{eV}$ respectively). The dashed line represents the performance of an endoreversible engine, a good approximation for the performance bound of TPV devices \cite{giteau_thermodynamic_2023}. All curves are obtained for $T_\mathrm{E} = 1700  \text{K}$ and $T_\mathrm{C} =  300 \text{K}$.}
    \label{fig:Figure 4}
\end{figure*}

\par{In Fig. \ref{fig:Figure 2}, we show the electrical power density and efficiency of a HCTPV system as a function of $\mu$ and $T$. We only consider positive values of $P_\mathrm{el}$, as negative power density corresponds to LED operation and is outside the scope of this paper. We consider $T_\mathrm{E} = 850  \text{K}$, $T_\mathrm{C} =  300 \text{K}$ and $E_\mathrm{g} = 0.74  \text{eV}$, corresponding to the bandgap of InGaAs lattice-matched to InP. The red curve in Fig. \ref{fig:Figure 2}(a) represents the analytically derived chemical potential $\mu$ that maximizes $P_\mathrm{el}$ for any temperature $T$ Eq. (\ref{eq: m_max}). This curve covers various different combinations of ($\mu$, $T$) that would all yield a similar $P_\mathrm{el}\approx 0.013 \ \mathrm{W.cm^{-2}}$ and efficiency $ \eta \approx 0.8 \ \eta_\mathrm{C}$. In Fig. \ref{fig:Figure 2}(b), we plot the corresponding HCTPV efficiency with respect to $\left(\mu, T\right)$. We confirm from Fig. \ref{fig:Figure 2}(b) that at open circuit voltage (that is, for $\mu = 0$ and $T = T_\mathrm{E}$), the efficiency reaches the Carnot limit (See Eq. (\ref{eq: eta_MB})).  We note that, unlike for conventional PV systems, the chemical potential can take negative values. Similarly, the carrier temperature can take values lower than $T_C$ or larger than $T_E$.} 

\par{Fig. \ref{fig:Figure 2} demonstrates the interplay between hot carrier temperature and chemical potential that defines HCTPV operation. For instance, a small hot carrier temperature only slightly above $T_\mathrm{C}$ would give negligible performance metrics in a standard TPV system. Nonetheless, in a HCTPV system, a large chemical potential $\mu$ can compensate for small $T$ and yield near-Carnot efficiency. Inversely, a very large hot carrier temperature does not necessarily improve power density as shown in Fig. \ref{fig:Figure 2} (a), as there is an optimal combination of $\mu$ and $T$ for $P_\mathrm{el}$ optimization. At the same time, from Fig. \ref{fig:Figure 2} (b) it is shown that, for all considered temperatures, one can approach Carnot efficiency by adjusting the chemical potential, a degree of freedom that is unavailable in standard TPV systems. We note, finally, that operating at low emitter temperature with a large bandgap strongly limits the benefits of hot carriers, because the cell may only absorb a small fraction of the radiation (see Fig. \ref{fig:Figure A3} in Appendix).} 

\par{In Fig. \ref{fig:Figure 2}, we described HCTPV operation in terms the chemical potential $\mu$ and temperature $T$ of the hot carriers. These quantities, however, cannot be controlled directly. Below, we describe the HCTPV system in terms of $E_\mathrm{ext}$ and $V$, which are externally controlled quantities of the device. Fig. \ref{fig:Figure 3} shows the figure of merits $P_\mathrm{el}$ and $\eta$, numerically calculated, as functions of $\left(V, E_\mathrm{ext}\right)$. For these calculations, we consider the same bandgap and lattice temperature, but double the emitter temperature ($T_\mathrm{E} = 1700 \ \text{K}$). In Fig. \ref{fig:Figure 3}(a), we demonstrate that maximum power density has improved by $500$ times with respect to the same device operating at half the emitter temperature (Fig. \ref{fig:Figure 2}(a)). This is expected as increasing $T_\mathrm{E}$ not only enhances total blackbody radiation by a factor $T_E^4$ but also increases the average energy of these photons, meaning a larger fraction of them are above the bandgap.} 

\par{From Fig. \ref{fig:Figure 3}(b), we confirm that the efficiency reaches the Carnot thermodynamic efficiency limit at open-circuit voltage for \textit{any} $E_\mathrm{ext}$, as per Eqs. (\ref{eq: voc_second_order_purelyhot})-(\ref{eq: efficiency_hot_voc}). By selecting $E_\mathrm{ext} = 1.27 E_\mathrm{g} \approx 0.94 \ \text{eV}$ such that the power density is maximized, the efficiency remains very high, nearly $0.8 \eta_\mathrm{C}$. We note that this is an important distinction between a standard TPV system and the HCTPV system considered here; having control over the energy level at which hot carriers are extracted is essentially the benefit of employing hot carriers in TPV systems. Finally, unlike for conventional PV systems, the voltage can be higher than the bandgap energy of the cell thanks to the temperature contribution of the hot carriers Eq. (\ref{eq: defn_hotcarrier_qV}).} 

\par{In a standard TPV system, optimal efficiency and optimal power density are achieved at significantly different voltages \cite{papadakis_broadening_2020}. By contrast, in Fig. \ref{fig:Figure 3} in combination with Eqs. (\ref{eq: voc_second_order_purelyhot}), we demonstrated that a HCTPV system reaches optimal power density at a voltage for which the efficiency penalty is very small.}

\par{In Fig. \ref{fig:Figure 4}, we vary the voltage applied to the HCTPV cell from $0$ to $V_\mathrm{oc}$ and plot the power-versus-efficiency performance curves. In Fig. \ref{fig:Figure 4}(a), we present the performance of the optimal HCTPV system considered above ($E_\mathrm{g} = 0.74 \ \mathrm{eV}$, $E_\mathrm{ext} = 0.94 \ \mathrm{eV}$). We compare this device to a conventional TPV cell with the same bandgap, for two distinct cases: broadband (blackbody) and narrowband thermal radiative exchange between the emitter and the cell. In the narrowband case, we consider a bandwidth of $0.1 k_\mathrm{B} T_\mathrm{E} \approx 15 \ \text{meV}$ at frequencies above the bandgap. First, we note that the power-efficiency trade-off between a broadband and a narrowband conventional TPV system can be clearly seen in this figure: a spectrally broadband TPV device reaches significantly higher power density, nonetheless this comes at the cost of a conversion efficiency that can never reach the Carnot limit. On the other hand, a spectrally narrowband TPV device does approach the Carnot limit, nonetheless its corresponding power density vanishes. In both cases, the power-versus-efficiency characteristic forms a loop, as the efficiency returns to $0$ at $V_\mathrm{oc}$.}

\par{By contrast, the HCTPV system not only yields significantly higher electrical power density than conventional broadband TPV system, it also reaches the Carnot limit at open-circuit voltage. As in the limiting cases of ultra-narrowband and broadband TPV systems, we note that the power-versus-efficiency characteristic curve of \textit{any} conventional ($T=T_\mathrm{C}$) TPV system will lie within the HCTPV curve. This means that an ideal HCTPV system will \textit{always} demonstrate an improved power-efficiency trade-off compared to a conventional, single-junction TPV one.}

\par{In Fig. \ref{fig:Figure 4}(b), we compare the performance of HCTPV systems to the thermodynamic bound of reciprocal radiative heat engines, approximated by the endoreversible model discussed in \cite{giteau_thermodynamic_2023}. This is described analytically through:
\begin{equation}
    \label{eq: endoreversible_limit}
    \rho = \eta \left[1 - \left(\frac{T_\mathrm{C}}{T_\mathrm{E}}\right)^4 \frac{1}{\left(1 - \eta\right)^4} \right].
\end{equation}
\noindent where $\rho$ is the electrical power density, $P_\mathrm{el}$, normalised to $\sigma T_\mathrm{E}^4$. 
Eq. (\ref{eq: endoreversible_limit}) is plotted in Fig. \ref{fig:Figure 4} (b) with the dashed black curve. We consider HCTPV systems with several bandgaps: 0.74 eV (InGaAs) 0.35 eV (InAs) and 0 eV, and plot their power-versus-efficiency characteristic curves at the extraction energy, $E_{ext}$, for which their power is optimized, respectively 0.94 eV, 0.59 eV and 0.41 eV. As shown, the power density \emph{for any efficiency} increases with decreasing bandgap as more hot-carriers contribute to the power density while no power gets dissipated.  As a limiting case, indeed, the zero-bandgap HCTPV system approaches the endoreversible model (Eq. (\ref{eq: endoreversible_limit})), black dashed curve), and even reaches a marginally higher maximum power point. This is because the endoreversible model considers a purely thermal engine with $\mu = 0$, while HCTPV systems offer control over both the temperature and the chemical potential, which can be leveraged to achieve slightly better performance.} 

\section{Conclusion}
\par{We highlight the potential benefits of harvesting hot carriers in TPV devices. Based on our theoretical formalism, we demonstrate that harnessing hot carriers alleviates the trade-off between electrical power output and conversion efficiency, allowing performance improvements in both metrics relative to conventional TPV cells. Specifically, we show that an ideal broadband hot-carrier TPV system can reach the Carnot efficiency limit at open circuit voltage, and that the power density at the optimal extraction energy is higher than that of conventional TPV devices, for any efficiency. We demonstrate that the power density substantially increases with the emitter temperature, whereas decreasing the bandgap consistently improves the performance.}

\par{In this work, we extend conclusions derived by solar PV community for the case of TPV systems. In particular, as demonstrated in the seminal HCSC analysis by Ross and Nozik \cite{ross_efficiency_1982}, the performance of a HCSC system is nearly identical to that of an infinite-junction SC. In other words the performance limit of a HCSC nearly overlaps with the multicolor limit. Analogously, in this work (Fig. \ref{fig:Figure 4}), we showed that a zero-bandgap hot-carrier TPV system is similar to an endoreversible radiative heat engine, approaching the thermodynamic performance bound of radiative heat engines. A multi-junction TPV system would have the same performance to a zero-bandgap HCTPV system, in the limit of infinite junctions.}

\par{As a final comment, we note that the results of this work can be further extended to near-field TPV operation, where an increased thermal radiative heat exchange can be exploited to saturate the thermalization channels \cite{le_bris_thermalisation_2012, nguyen_quantitative_2018,giteau_detailed_2019}. Such a synergy between near-field radiative heat transfer and hot carriers may enable substantial improvement in realistic TPV devices.}

\begin{acknowledgments}
\par{This work is dedicated to the memory of John S. Papadakis. The authors declare no competing financial interest. G.T.P. acknowledges financial support from the la Caixa Foundation (ID 00010434). M. G. acknowledges financial support from the Severo Ochoa Excellence Fellowship. This work was supported by the Spanish MICINN (PID2021-125441OA-I00, PID2020–112625GB-I00, and CEX2019-000910- S), the European Union (fellowship LCF/BQ/PI21/11830019 under the Marie Skłodowska-Curie Grant Agreement No. 847648), Generalitat de Catalunya (2021 SGR 01443) through the CERCA program, Fundació Cellex, and Fundació Mir-Puig.}
\end{acknowledgments}

\section*{Appendix}
\setcounter{equation}{0}
\setcounter{figure}{0}
\renewcommand{\thefigure}{A\arabic{figure}}
\renewcommand{\theequation}{A\arabic{equation}} 

\subsection{Analytics in the Maxwell-Boltzmann approximation}
\par{Here, we derive the analytical expressions of the power density and the efficiency using Eq. (\ref{eq: defn_j_abs})-(\ref{eq: defn_j_rad}) in the main text. We consider the Maxwell-Boltzmann (MB) approximation, namely, we approximate the occupation number of photons by:
\begin{equation}
    \frac{1}{e^{(\varepsilon-\mu) / k_B T}-1}  \approx e^{\mu - \varepsilon/k_B T}.
\end{equation}}

\par{For the sake of algebraic simplicity, we normalize all the parameters to the bandgap energy of the semiconductor:
\begin{align*}
    m &= \dfrac{\mu}{E_\mathrm{g}},
    \\
    t &= \dfrac{k_B T}{E_\mathrm{g}},
    \\
    \tau_E &= \dfrac{k_B T_E}{E_\mathrm{g}},
    \\
    \tau_C &= \dfrac{k_B T_C}{E_\mathrm{g}}.
\end{align*}
Retaining terms up to the second order in $t$, we obtain the following expressions for radiated particle flux and power density:
\begin{align}
    \label{eq: second_Bapprox_j_farfield}
    N^\mathrm{MB}_\mathrm{rad}\left(m,t \right) &= \frac{E_\mathrm{g}^3}{4 \mathrm{c}^2 \pi ^2 \hbar^3} \ t \ \left(1 + 2 t\right) \ e^{\frac{m-1}{t}}
    \\
    \label{eq: second_Bapprox_p_farfield}
    P^\mathrm{MB}_\mathrm{rad} \left(m,t\right) &=  \frac{E_\mathrm{g}^4}{4 \mathrm{c}^2 \pi ^2 \hbar^3} \ t \ \left(1 + 3 t\right) \ e^{\frac{m-1}{t}}.
\end{align}}

\par{The absorbed fluxes, $N^{\mathrm{MB}}_\mathrm{abs}$ and $P^{\mathrm{MB}}_\mathrm{abs}$, are obtained similarly by replacing $\left(m, t\right)$ with $\left(0, \tau_\mathrm{E}\right)$ in Eq. (\ref{eq: second_Bapprox_j_farfield})-(\ref{eq: second_Bapprox_p_farfield}). From here, we have an analytical expression for the electrical power density as a function of the internal parameters $m$ and $t$, using Eq. (\ref{eq: defn_hotcarrier_qV}) in the main text:
\begin{equation}
    \label{eq: defn_hotcarrier_pout}
        P_\mathrm{el}^{\mathrm{MB}} = \frac{\tau_\mathrm{C} }{t} m E_\mathrm{g} N_\mathrm{ext}^{\mathrm{MB}} +  \left(1-\frac{\tau_\mathrm{C}}{t}\right) P_\mathrm{ext}^{\mathrm{MB}},
\end{equation}
\noindent where 
\begin{align}
    \label{eq: second_Bapprox_jext_farfield}
    N_\mathrm{ext}^{\mathrm{MB}} &= \frac{E_\mathrm{g}^3}{4 \mathrm{c}^2 \pi ^2 \hbar^3} \left[ \tau_E \left(1 + 2\tau_E\right) e^{-\frac{1}{\tau_E}} - t \left(1 + 2t\right) e^{\frac{m - 1}{t}} \right],
    \\
    \label{eq: second_Bapprox_pext_farfield}
    P_\mathrm{ext}^{\mathrm{MB}} &=  \frac{E_\mathrm{g}^4}{4 \mathrm{c}^2 \pi ^2 \hbar^3} \left[ \tau_E\left(1 + 3 \tau_E \right) e^{-\frac{1}{\tau_E}} - t\left(1 + 3 t\right) e^{\frac{m - 1}{t}} \right].
\end{align}}

\begin{figure}
    \includegraphics[width=\linewidth]{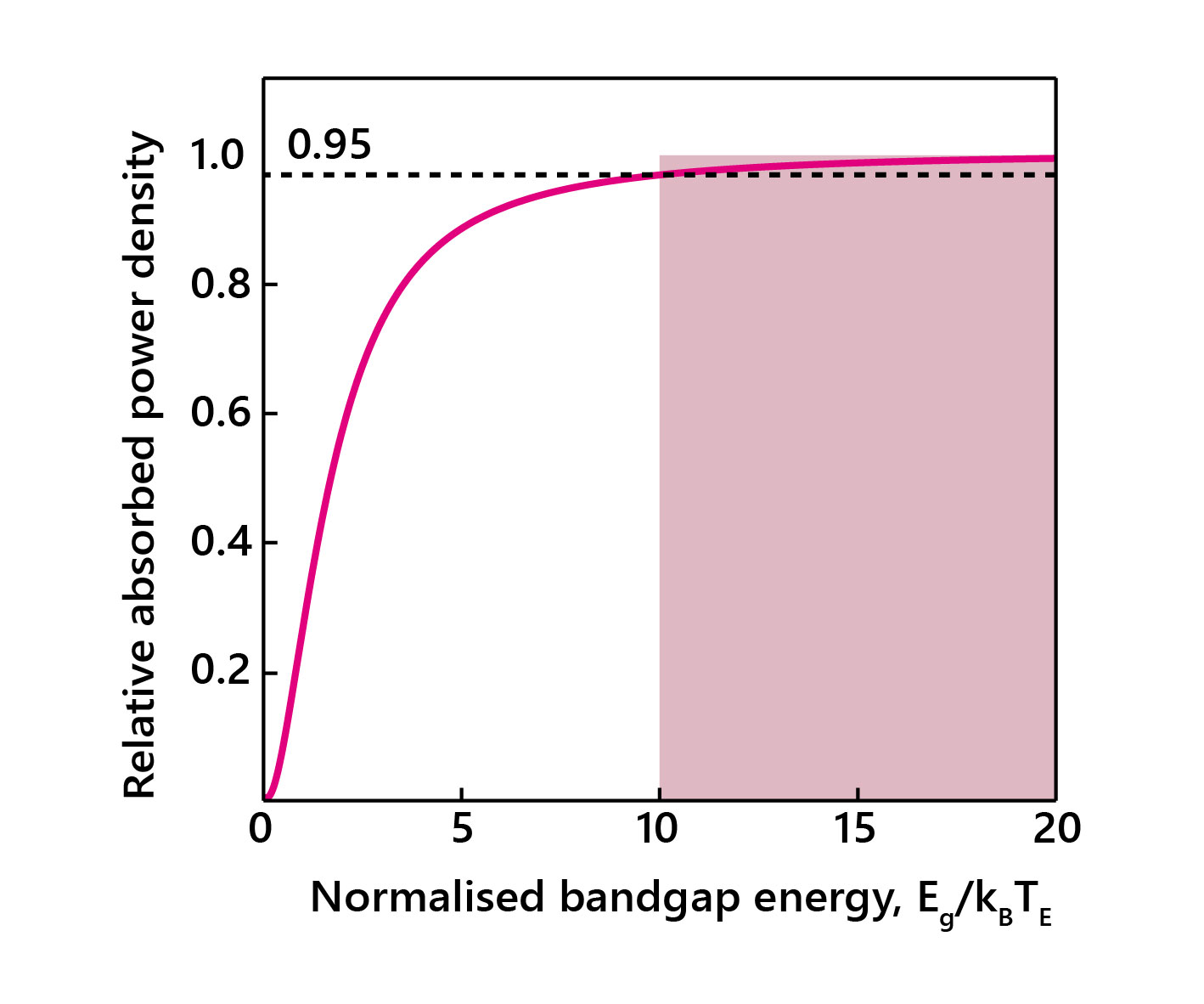}
    \caption{Absorbed power density ($P_{abs}^\mathrm{MB}$) normalized to $\sigma T_\mathrm{E}^{4}$ versus bandgap energy in terms of $k_\mathrm{B}T_\mathrm{E}$. The second-order Maxwell-Boltzmann approximation is valid above 95$\%$ for $E_\mathrm{g}/k_\mathrm{B}T_\mathrm{E} \geq 10$. For $E_\mathrm{g} = 0.74  \text{eV}$ and $T_\mathrm{E} = 850 \text{K}$, $E_\mathrm{g}/k_\mathrm{B}T_\mathrm{E} \approx 10$.}
    \label{fig:Figure A1}
\end{figure}

\par{This allows us to write the extraction energy as:
\begin{align}
    \label{eq: extraction_energy}
    E^\mathrm{MB}_\mathrm{{ext}} = E_\mathrm{g} \left[ 1 + \Theta\left(m, t\right)\right],
\end{align}
\noindent where
\begin{align}
    \Theta(m,t) = \frac{\tau_\mathrm{E}^2 e^{-\frac{1}{\tau_\mathrm{E}}} - t^2 e^{\frac{m - 1}{t}}}{\tau_\mathrm{E} \left(1 + 2 \tau_\mathrm{E} \right) e^{-\frac{1}{\tau_\mathrm{E}}} - t \left(1 + 2t \right) e^{\frac{m-1}{t}}}.
\end{align}}

\begin{figure*}
\includegraphics[width=\linewidth]{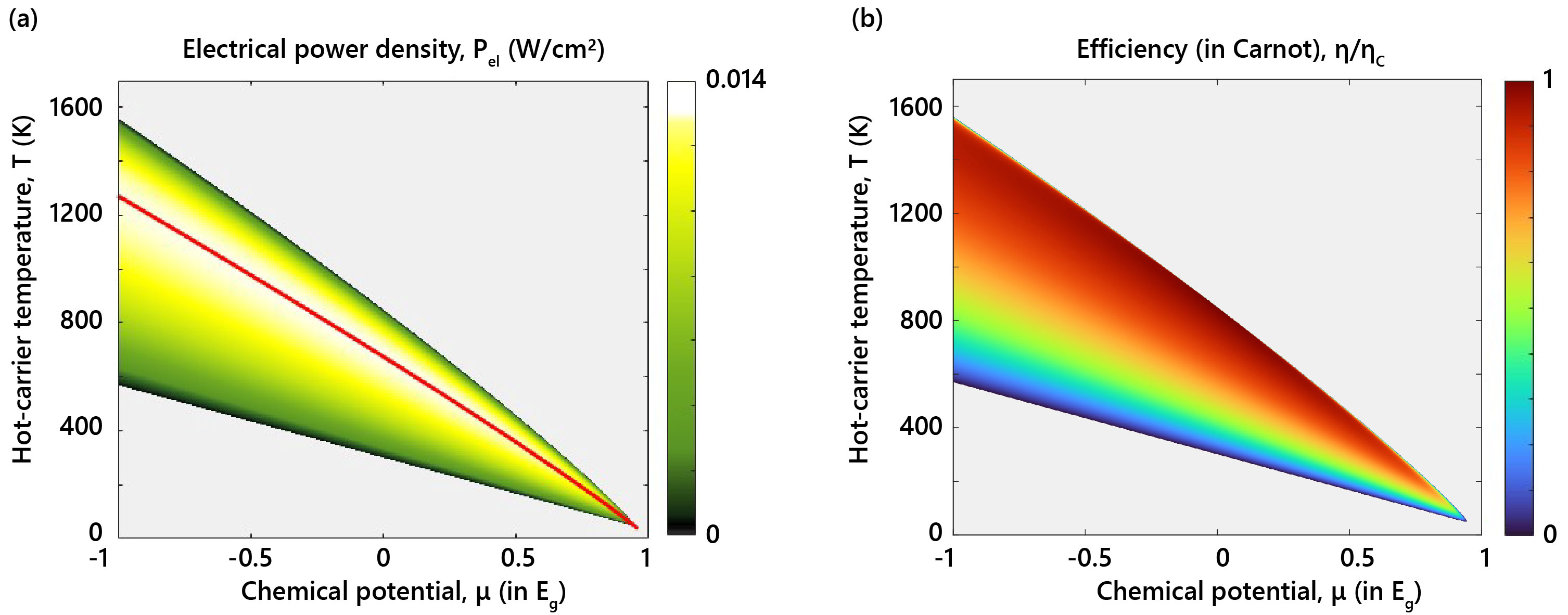}
\caption{(a) Electrical output power density and, (b) efficiency numerically calculated as a function of normalised chemical potential and temperature of the hot-carriers at $T_\mathrm{E} = 850  \text{K}$, $T_\mathrm{C} =  300 \text{K}$ and $E_\mathrm{g} = 0.74  \text{eV}$. The red curve in (a) represents the temperature as a function of chemical potential for which the power density is maximum, given by Eq. (\ref{eq: m_max}). The gray regions correspond to $P_\mathrm{el} <0$ (not relevant for TPV operation).}
\label{fig:Figure A2}
\end{figure*}

\par{Furthermore, we observed in Fig. 2(a) (in the main text) that the power density reaches its maximum value for particular values of $\left(\mu, T\right)$, represented with a red line. Here, we derive the chemical potential $m_0$ at which the power density is maximized for a given $t$ by setting the derivative of $P_{\mathrm{el}}^{\mathrm{MB}}$ with respect to $m$ equal to zero. We find:
\begin{equation}
\label{eq: m_max}
m_\mathrm{0} = \frac{\tau_\mathrm{C} -t \left[2 (t-1) \tau_\mathrm{C}+3 t+1 \right]}{(2t+1) \tau_\mathrm{C}}+t \ \mathrm{W}\left[\alpha\right],
\end{equation}
\noindent where $\mathrm{W}$ is the Lambert W function and
\begin{equation}
    \alpha = \frac{\tau_\mathrm{E}(2 \tau_\mathrm{E}+1)}{t(2t+1)} \exp\left({\frac{2 t \tau_\mathrm{E} \tau_\mathrm{C}+3 t \tau_\mathrm{E}-2 t \tau_\mathrm{C}+\tau_\mathrm{E}-\tau_\mathrm{C}}{2 t \tau_\mathrm{E} \tau_\mathrm{C}+\tau_\mathrm{E}\tau_\mathrm{C}}}\right).
\end{equation}}
\par{Finally, we derive the efficiency of the hot-carrier TPV system in the MB approximation using Eq. (\ref{eq: defn_hotcarrier_qV})-(\ref{eq: defn_efficiency_hot}), from the main text and Eq. (\ref{eq: extraction_energy}), leading to:
\begin{equation}
    \label{eq: eta_MB}
  \eta^{\mathrm{MB}} = 1 - \dfrac{\tau_C}{t} \left(1 - \dfrac{m}{1 + \Theta\left(m, t \right)}\right). 
\end{equation}
At open circuit voltage, that is, for $m = 0$ and $t = \tau_\mathrm{E}$, we recover $\eta^{\mathrm{MB}} = \eta_\mathrm{C}$.}

\subsection{Validity of the MB approximation}
\par{We compare the power density obtained using the MB approximation, $P_{abs}^{\mathrm{MB}}$, with the exact power density given by integrating Planck's law from $\omega_\mathrm{g}$ to $\infty$. In Fig. \ref{fig:Figure A1}, we plot the relative absorbed power as a function of $E_\mathrm{g}/k_\mathrm{B}T_\mathrm{E}$. We observe that the power absorbed using the MB approximation reaches 95$\%$ of its actual value for $E_\mathrm{g}/k_\mathrm{B}T_\mathrm{E} \geq 10$, implying that the second-order MB approximation holds true within the shaded region.} 

\par{Furthermore, the validity of the MB approximation in the domain $E_\mathrm{g}/k_\mathrm{B}T_\mathrm{E} \approx 10$ can be confirmed in Fig. \ref{fig:Figure A2}, where the exact numerical results closely match the analytical results presented in Fig. \ref{fig:Figure 2} of the main text. The red curve, derived analytically using the Maxwell-Boltzmann approximation (Eq. (\ref{eq: m_max})), coincides with the region indicating the maximum electrical power density.}

\begin{figure}[ht!]
    \includegraphics[width=\linewidth]{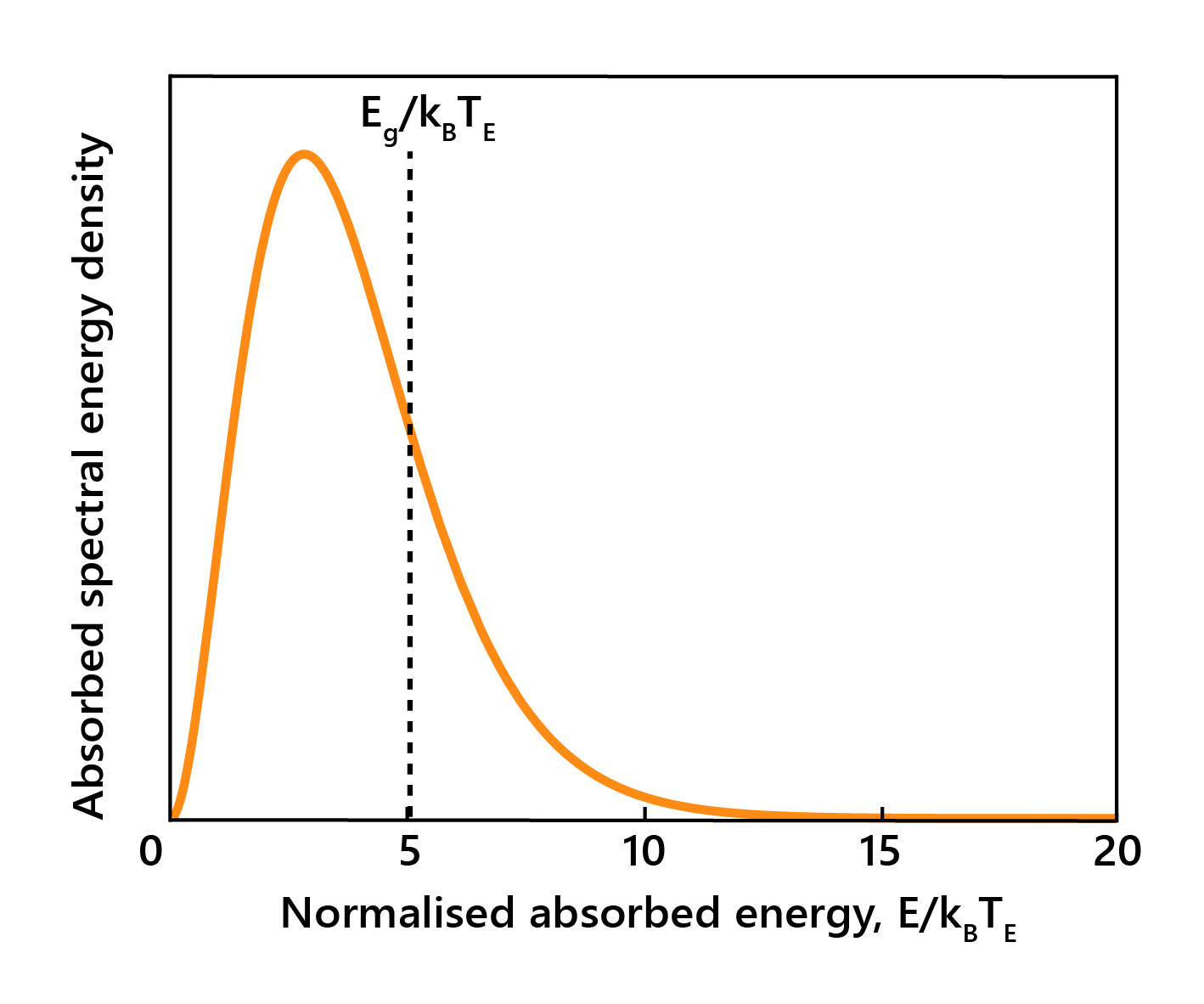}
    \caption{Absorbed spectral energy density as a function of normalized absorbed energy ($E/k_\mathrm{B} T_\mathrm{E}$). The dashed line indicates $E_\mathrm{g}/k_\mathrm{B} T_\mathrm{E}$ for $E_\mathrm{g} = 0.74  \text{eV}$ and $T_\mathrm{E} = 1700 \text{K}$.}
    \label{fig:Figure A3}
\end{figure}

\par{Finally, we show in Fig. \ref{fig:Figure A3} the spectral energy density radiated by the emitter as a function of the photon energy (normalized to $k_\mathrm{B}T_\mathrm{E}$). The absorbed particle flux (Eq. (\ref{eq: defn_j_abs})) and power density is obtained by integrating this quantity from $E_\mathrm{g}/k_\mathrm{B}T_\mathrm{E}$ to $\infty$. We note that being within the validity domain of the MB approximation ($E_\mathrm{g}/k_\mathrm{B}T_\mathrm{E} \geq 10$, see Fig. \ref{fig:Figure A1}) implies a very small portion of the spectrum is absorbed in the cell, strongly limiting the power output. As a result, larger values of $E_\mathrm{g}/k_\mathrm{B}T_\mathrm{E}$ restrict the potential benefits of harnessing hot-carriers, as only the high-energy tail of the incident spectrum can be absorbed. In fact, raising $T_\mathrm{E}$ not only shifts the dashed line (represented by the dashed line in Fig. \ref{fig:Figure A3}) to the left, but also increases blackbody radiation, resulting in higher power density. Thus, low-bandgap semiconductors and high emitter temperatures synergistically contribute to the improved performance of hot-carrier TPV systems. As a result, high-power HC-TPV is not compatible with the MB approximation.}

\bibliographystyle{ieeetr}
\bibliography{references.bib}

\begin{thebibliography}{10}

\bibitem{harder_theoretical_2003}
N.-P. Harder and P.~Würfel, ``Theoretical limits of thermophotovoltaic solar energy conversion,'' {\em Semiconductor Science and Technology}, vol.~18, pp.~S151--S157, May 2003.

\bibitem{chubb_fundamentals_2007}
D.~L. Chubb, {\em Fundamentals of thermophotovoltaic energy conversion}.
\newblock Amsterdam, Netherlands ; Boston: Elsevier, 1st ed~ed., 2007.
\newblock OCLC: ocm84150882.

\bibitem{datas_steady_2013}
A.~Datas, D.~Chubb, and A.~Veeraragavan, ``Steady state analysis of a storage integrated solar thermophotovoltaic ({SISTPV}) system,'' {\em Solar Energy}, vol.~96, pp.~33--45, Oct. 2013.

\bibitem{karalis_squeezing_2016}
A.~Karalis and J.~D. Joannopoulos, ``‘{Squeezing}’ near-field thermal emission for ultra-efficient high-power thermophotovoltaic conversion,'' {\em Scientific Reports}, vol.~6, p.~28472, July 2016.

\bibitem{papadakis_thermodynamics_2021}
G.~T. Papadakis, M.~Orenstein, E.~Yablonovitch, and S.~Fan, ``Thermodynamics of {Light} {Management} in {Near}-{Field} {Thermophotovoltaics},'' {\em Physical Review Applied}, vol.~16, p.~064063, Dec. 2021.

\bibitem{lapotin_thermophotovoltaic_2022}
A.~LaPotin, K.~L. Schulte, M.~A. Steiner, K.~Buznitsky, C.~C. Kelsall, D.~J. Friedman, E.~J. Tervo, R.~M. France, M.~R. Young, A.~Rohskopf, S.~Verma, E.~N. Wang, and A.~Henry, ``Thermophotovoltaic efficiency of 40\%,'' {\em Nature}, vol.~604, pp.~287--291, Apr. 2022.

\bibitem{omair_ultraefficient_2019}
Z.~Omair, G.~Scranton, L.~M. Pazos-Outón, T.~P. Xiao, M.~A. Steiner, V.~Ganapati, P.~F. Peterson, J.~Holzrichter, H.~Atwater, and E.~Yablonovitch, ``Ultraefficient thermophotovoltaic power conversion by band-edge spectral filtering,'' {\em Proceedings of the National Academy of Sciences}, vol.~116, pp.~15356--15361, July 2019.

\bibitem{datas_thermophotovoltaic_2021}
A.~Datas and R.~Vaillon, ``Thermophotovoltaic energy conversion,'' in {\em Ultra-{High} {Temperature} {Thermal} {Energy} {Storage}, {Transfer} and {Conversion}}, pp.~285--308, Elsevier, 2021.

\bibitem{zhao_high-performance_2017}
B.~Zhao, K.~Chen, S.~Buddhiraju, G.~Bhatt, M.~Lipson, and S.~Fan, ``High-performance near-field thermophotovoltaics for waste heat recovery,'' {\em Nano Energy}, vol.~41, pp.~344--350, Nov. 2017.

\bibitem{giteau_thermodynamic_2023}
M.~Giteau, M.~F. Picardi, and G.~T. Papadakis, ``Thermodynamic performance bounds for radiative heat engines,'' July 2023.
\newblock arXiv:2304.03942 [physics].

\bibitem{tervo_efficient_2022}
E.~J. Tervo, R.~M. France, D.~J. Friedman, M.~K. Arulanandam, R.~R. King, T.~C. Narayan, C.~Luciano, D.~P. Nizamian, B.~A. Johnson, A.~R. Young, L.~Y. Kuritzky, E.~E. Perl, M.~Limpinsel, B.~M. Kayes, A.~J. Ponec, D.~M. Bierman, J.~A. Briggs, and M.~A. Steiner, ``Efficient and scalable {GaInAs} thermophotovoltaic devices,'' {\em Joule}, vol.~6, pp.~2566--2584, Nov. 2022.

\bibitem{mittapally_near-field_2021}
R.~Mittapally, B.~Lee, L.~Zhu, A.~Reihani, J.~W. Lim, D.~Fan, S.~R. Forrest, P.~Reddy, and E.~Meyhofer, ``Near-field thermophotovoltaics for efficient heat to electricity conversion at high power density,'' {\em Nature Communications}, vol.~12, p.~4364, July 2021.

\bibitem{song_modeling_2022}
J.~Song, J.~Han, M.~Choi, and B.~J. Lee, ``Modeling and experiments of near-field thermophotovoltaic conversion: {A} review,'' {\em Solar Energy Materials and Solar Cells}, vol.~238, p.~111556, May 2022.

\bibitem{fan_near-perfect_2020}
D.~Fan, T.~Burger, S.~McSherry, B.~Lee, A.~Lenert, and S.~R. Forrest, ``Near-perfect photon utilization in an air-bridge thermophotovoltaic cell,'' {\em Nature}, vol.~586, pp.~237--241, Oct. 2020.

\bibitem{lee_air-bridge_2022}
B.~Lee, R.~Lentz, T.~Burger, B.~Roy-Layinde, J.~Lim, R.~M. Zhu, D.~Fan, A.~Lenert, and S.~R. Forrest, ``Air-{Bridge} {Si} {Thermophotovoltaic} {Cell} with {High} {Photon} {Utilization},'' {\em ACS Energy Letters}, vol.~7, pp.~2388--2392, July 2022.

\bibitem{lim_enhanced_2023}
J.~Lim, B.~Roy-Layinde, B.~Liu, A.~Lenert, and S.~R. Forrest, ``Enhanced {Photon} {Utilization} in {Single} {Cavity} {Mode} {Air}-{Bridge} {Thermophotovoltaic} {Cells},'' {\em ACS Energy Letters}, pp.~2935--2939, June 2023.

\bibitem{lopez_thermophotovoltaic_2023}
E.~López, I.~Artacho, and A.~Datas, ``Thermophotovoltaic conversion efficiency measurement at high view factors,'' {\em Solar Energy Materials and Solar Cells}, vol.~250, p.~112069, Jan. 2023.

\bibitem{pendry_radiative_1999}
J.~B. Pendry, ``Radiative exchange of heat between nanostructures,'' {\em Journal of Physics: Condensed Matter}, vol.~11, pp.~6621--6633, Sept. 1999.

\bibitem{lucchesi_near-field_2021}
C.~Lucchesi, D.~Cakiroglu, J.-P. Perez, T.~Taliercio, E.~Tournié, P.-O. Chapuis, and R.~Vaillon, ``Near-{Field} {Thermophotovoltaic} {Conversion} with {High} {Electrical} {Power} {Density} and {Cell} {Efficiency} above 14\%,'' {\em Nano Letters}, vol.~21, pp.~4524--4529, June 2021.

\bibitem{pascale_perspective_2023}
M.~Pascale, M.~Giteau, and G.~T. Papadakis, ``Perspective on near-field radiative heat transfer,'' {\em Applied Physics Letters}, vol.~122, p.~100501, Mar. 2023.

\bibitem{shockley_detailed_2004}
W.~Shockley and H.~J. Queisser, ``Detailed {Balance} {Limit} of {Efficiency} of p‐n {Junction} {Solar} {Cells},'' {\em Journal of Applied Physics}, vol.~32, pp.~510--519, June 2004.

\bibitem{green_third_2006}
M.~A. Green, {\em Third generation photovoltaics: advanced solar energy conversion}.
\newblock No.~12 in Springer series in photonics, Berlin ; New York: Springer, 2006.
\newblock OCLC: ocm63144200.

\bibitem{vos_detailed_1980}
A.~D. Vos, ``Detailed balance limit of the efficiency of tandem solar cells,'' {\em Journal of Physics D: Applied Physics}, vol.~13, pp.~839--846, May 1980.

\bibitem{geisz_six-junction_2020}
J.~F. Geisz, R.~M. France, K.~L. Schulte, M.~A. Steiner, A.~G. Norman, H.~L. Guthrey, M.~R. Young, T.~Song, and T.~Moriarty, ``Six-junction {III}–{V} solar cells with 47.1\% conversion efficiency under 143 {Suns} concentration,'' {\em Nature Energy}, vol.~5, pp.~326--335, Apr. 2020.

\bibitem{green_solar_2023}
M.~A. Green, E.~D. Dunlop, G.~Siefer, M.~Yoshita, N.~Kopidakis, K.~Bothe, and X.~Hao, ``Solar cell efficiency tables ({Version} 61),'' {\em Progress in Photovoltaics: Research and Applications}, vol.~31, pp.~3--16, Jan. 2023.

\bibitem{datas_optimum_2015}
A.~Datas, ``Optimum semiconductor bandgaps in single junction and multijunction thermophotovoltaic converters,'' {\em Solar Energy Materials and Solar Cells}, vol.~134, pp.~275--290, Mar. 2015.

\bibitem{ross_efficiency_1982}
R.~T. Ross and A.~J. Nozik, ``Efficiency of hot‐carrier solar energy converters,'' {\em Journal of Applied Physics}, vol.~53, pp.~3813--3818, May 1982.

\bibitem{wurfel_solar_1997}
P.~Würfel, ``Solar energy conversion with hot electrons from impact ionisation,'' {\em Solar Energy Materials and Solar Cells}, vol.~46, pp.~43--52, Apr. 1997.

\bibitem{marti_thermodynamics_2022}
A.~Martí, E.~Antolín, and I.~Ramiro, ``Thermodynamics of the {Monoenergetic} {Energy}-{Selective} {Contacts} of a {Hot}-{Carrier} {Solar} {Cell},'' {\em Physical Review Applied}, vol.~18, p.~064048, Dec. 2022.

\bibitem{konig_non-equilibrium_2020}
D.~König, Y.~Yao, B.~Puthen-Veettil, and S.~C. Smith, ``Non-equilibrium dynamics, materials and structures for hot carrier solar cells: a detailed review,'' {\em Semiconductor Science and Technology}, vol.~35, p.~073002, July 2020.

\bibitem{zhang_review_2021}
Y.~Zhang, X.~Jia, S.~Liu, B.~Zhang, K.~Lin, J.~Zhang, and G.~Conibeer, ``A review on thermalization mechanisms and prospect absorber materials for the hot carrier solar cells,'' {\em Solar Energy Materials and Solar Cells}, vol.~225, p.~111073, June 2021.

\bibitem{giteau_identification_2020}
M.~Giteau, E.~de~Moustier, D.~Suchet, H.~Esmaielpour, H.~Sodabanlu, K.~Watanabe, S.~Collin, J.-F. Guillemoles, and Y.~Okada, ``Identification of surface and volume hot-carrier thermalization mechanisms in ultrathin {GaAs} layers,'' {\em Journal of Applied Physics}, vol.~128, p.~193102, Nov. 2020.

\bibitem{rosenwaks_hot-carrier_1993}
Y.~Rosenwaks, M.~C. Hanna, D.~H. Levi, D.~M. Szmyd, R.~K. Ahrenkiel, and A.~J. Nozik, ``Hot-carrier cooling in {GaAs}: {Quantum} wells versus bulk,'' {\em Physical Review B}, vol.~48, pp.~14675--14678, Nov. 1993.

\bibitem{hirst_enhanced_2014}
L.~C. Hirst, M.~K. Yakes, C.~G. Bailey, J.~G. Tischler, M.~P. Lumb, M.~Gonzalez, M.~F. Fuhrer, N.~J. Ekins-Daukes, and R.~J. Walters, ``Enhanced {Hot}-{Carrier} {Effects} in {InAlAs}/{InGaAs} {Quantum} {Wells},'' {\em IEEE Journal of Photovoltaics}, vol.~4, pp.~1526--1531, Nov. 2014.

\bibitem{nguyen_quantitative_2018}
D.-T. Nguyen, L.~Lombez, F.~Gibelli, S.~Boyer-Richard, A.~Le~Corre, O.~Durand, and J.-F. Guillemoles, ``Quantitative experimental assessment of hot carrier-enhanced solar cells at room temperature,'' {\em Nature Energy}, vol.~3, pp.~236--242, Mar. 2018.

\bibitem{makhfudz_enhancement_2022}
I.~Makhfudz, N.~Cavassilas, M.~Giteau, H.~Esmaielpour, D.~Suchet, A.-M. Daré, and F.~Michelini, ``Enhancement of hot carrier effect and signatures of confinement in terms of thermalization power in quantum well solar cell,'' {\em Journal of Physics D: Applied Physics}, vol.~55, p.~475102, Oct. 2022.
\newblock Publisher: IOP Publishing.

\bibitem{fast_hot-carrier_2021}
J.~Fast, U.~Aeberhard, S.~P. Bremner, and H.~Linke, ``Hot-carrier optoelectronic devices based on semiconductor nanowires,'' {\em Applied Physics Reviews}, vol.~8, p.~021309, Apr. 2021.

\bibitem{harada_hot-carrier_2019}
Y.~Harada, N.~Iwata, S.~Asahi, and T.~Kita, ``Hot-carrier generation and extraction in {InAs}/{GaAs} quantum dot superlattice solar cells,'' {\em Semiconductor Science and Technology}, vol.~34, p.~094003, Aug. 2019.
\newblock Publisher: IOP Publishing.

\bibitem{giteau_hot-carrier_2022-1}
M.~Giteau, S.~Almosni, J.-F. Guillemoles, and D.~Suchet, ``Hot-carrier multijunction solar cells: sensitivity and resilience to nonidealities,'' {\em Journal of Photonics for Energy}, vol.~12, p.~032208, June 2022.
\newblock Publisher: SPIE.

\bibitem{giteau_hot-carrier_2022}
M.~Giteau, S.~Almosni, and J.-F. Guillemoles, ``Hot-carrier multi-junction solar cells: {A} synergistic approach,'' {\em Applied Physics Letters}, vol.~120, p.~213901, May 2022.

\bibitem{st-gelais_hot_2017}
R.~St-Gelais, G.~R. Bhatt, L.~Zhu, S.~Fan, and M.~Lipson, ``Hot {Carrier}-{Based} {Near}-{Field} {Thermophotovoltaic} {Energy} {Conversion},'' {\em ACS Nano}, vol.~11, pp.~3001--3009, Mar. 2017.

\bibitem{wang_hot_2023}
J.~Wang, Y.~Wang, X.~Chen, J.~Chen, and S.~Su, ``Hot carrier-based near-field thermophotovoltaics with energy selective contacts,'' {\em Applied Physics Letters}, vol.~122, p.~122203, Mar. 2023.

\bibitem{burger_present_2020}
T.~Burger, C.~Sempere, B.~Roy-Layinde, and A.~Lenert, ``Present {Efficiencies} and {Future} {Opportunities} in {Thermophotovoltaics},'' {\em Joule}, vol.~4, pp.~1660--1680, Aug. 2020.

\bibitem{wurfel_chemical_1982}
P.~Würfel, ``The chemical potential of radiation,'' {\em Journal of Physics C: Solid State Physics}, vol.~15, pp.~3967--3985, June 1982.

\bibitem{giteau_detailed_2019}
M.~Giteau, D.~Suchet, S.~Collin, J.-F. Guillemoles, and Y.~Okada, ``Detailed balance calculations for hot-carrier solar cells: coupling high absorptivity with low thermalization through light trapping,'' {\em EPJ Photovoltaics}, vol.~10, p.~1, 2019.

\bibitem{papadakis_broadening_2020}
G.~T. Papadakis, S.~Buddhiraju, Z.~Zhao, B.~Zhao, and S.~Fan, ``Broadening {Near}-{Field} {Emission} for {Performance} {Enhancement} in {Thermophotovoltaics},'' {\em Nano Letters}, vol.~20, pp.~1654--1661, Mar. 2020.

\bibitem{le_bris_thermalisation_2012}
A.~Le~Bris, L.~Lombez, S.~Laribi, G.~Boissier, P.~Christol, and J.-F. Guillemoles, ``Thermalisation rate study of {GaSb}-based heterostructures by continuous wave photoluminescence and their potential as hot carrier solar cell absorbers,'' {\em Energy \& Environmental Science}, vol.~5, no.~3, p.~6225, 2012.

\end{thebibliography}

\end{document}